\documentclass[apj]{emulateapj}
\slugcomment{{\sc Accepted to ApJL:} March 22, 2010}

\usepackage{graphicx}
\usepackage{epstopdf}
\usepackage{stmaryrd}
\usepackage{amsmath}
\usepackage{ccaption}
\usepackage{multirow}
\usepackage{bm}
\usepackage{nicefrac}
\usepackage{comment}
\usepackage{color}

\newcommand{\bdm}{\begin{displaymath}}
\newcommand{\edm}{\end{displaymath}}
\newcommand{\beq}{\begin{equation}}
\newcommand{\eeq}{\end{equation}}
\newcommand{\bit}{\begin{itemize}}
\newcommand{\eit}{\end{itemize}}
\newcommand{\ben}{\begin{enumerate}}
\newcommand{\een}{\end{enumerate}}
\newcommand{\bfi}{\begin{figure}[htb]}
\newcommand{\bpfi}{\begin{figure}[p]}

\shorttitle{IR-bright galaxies on the IR-radio relation}
\shortauthors{Sargent et al.}

\begin{document}


\title{No Evolution in the IR-Radio Relation for IR-Luminous Galaxies at z\,$<$\,2 in the COSMOS Field\altaffilmark{$\dagger$}}

\author{M.~T. Sargent\altaffilmark{1, $\star$},
E. Schinnerer\altaffilmark{1},
E. Murphy\altaffilmark{2},
C.~L. Carilli\altaffilmark{3},
G. Helou\altaffilmark{4},
H. Aussel\altaffilmark{5},
E. Le Floc'h\altaffilmark{5},
D.~T. Frayer\altaffilmark{6},
O. Ilbert\altaffilmark{7},
P. Oesch\altaffilmark{8},
M. Salvato\altaffilmark{9},
V. Smol\v{c}i\'{c}\altaffilmark{10},
J. Kartaltepe\altaffilmark{11},
D.~B. Sanders\altaffilmark{12}
}

\altaffiltext{$\star$}{E-mail: \texttt{markmr@mpia.de}}

\altaffiltext{1}{Max-Planck-Institut f\"ur Astronomie, K\"onigstuhl 17,
                 D-69117 Heidelberg, Germany}
\altaffiltext{2}{{\it Spitzer} Science Center, MC 314-6, California Institute of Technology,
                 Pasadena, CA 91125}
\altaffiltext{3}{National Radio Astronomy Observatory, P.O. Box 0, Socorro,
                 NM 87801-0387, USA}
\altaffiltext{4}{Infrared Processing and Analysis Center, MC 100-22, California Institute of Technology,
                 Pasadena, CA 91125}
\altaffiltext{5}{AIM Unit\'e Mixte de Recherche CEA CNRS Universit\'e
                 Paris VII UMR n158, France}
\altaffiltext{6}{National Radio Astronomy Observatory,
                 P.O. Box 2, Green Bank, WV 24944, USA}
\altaffiltext{7}{Laboratoire d'Astrophysique de Marseille, Universit\'e de Provence, CNRS, 38 rue Fr\'ed\'eric Joliot-Curie,
                 F-13388 Marseille Cedex 13, France}
\altaffiltext{8}{Department of Physics, ETH Zurich,
                 CH-8093 Zurich, Switzerland}
\altaffiltext{9}{Max-Planck-Institut f\"ur Plasmaphysik, Boltzmanstrasse,
                 D-85741 Garching, Germany}
\altaffiltext{10}{California Institute of Technology, MC 105-24, 1200 East
                 California Boulevard, Pasadena, CA 91125, USA}
\altaffiltext{11}{National Optical Astronomy Observatory,
                 950 N. Cherry Ave, Tucson, AZ 85726, USA}

\altaffiltext{12}{Institute for Astronomy, 2680 Woodlawn Dr., University
                 of Hawaii, Honolulu, Hawaii, 96822, USA}
%
%
%
%
\altaffiltext{$\dagger$}{Partly based on observations collected at the European Organisation for Astronomical Research in the Southern Hemisphere, Chile, ESO program ID 175.A-0839.}

\begin{abstract}
Previous observational studies of the infrared (IR)-radio relation out to high redshift employed any detectable star forming systems at a given redshift within the restricted area of cosmological survey fields. Consequently, the evolution inferred relies on a comparison between the average IR/radio properties of ({\it i}\,) very IR-luminous high-$z$ sources and ({\it ii}\,) more heterogeneous low(er)-$z$ samples that often lack the strongest IR emitters. In this report we consider populations of objects with comparable luminosities over the last 10\,Gyr by taking advantage of deep IR (esp. {\it Spitzer} 24\,$\mu$m) and VLA 1.4\,GHz observations of the COSMOS field. Consistent with recent model predictions, both Ultra Luminous Infrared Galaxies (ULIRGs) and galaxies on the bright end of the evolving IR luminosity function do not display any change in their average IR/radio ratios out to $z$\,$\sim$\,2 when corrected for bias. Uncorrected data suggested $\sim$0.3 dex of positive evolution.
\end{abstract}


\section{Introduction}

The IR/radio properties of galaxies at successively higher redshift have been probed in the past decade using either statistical samples from cosmological survey fields \citep[e.g.][hereafter: S10]{appleton04, frayer06, sargent10}, the  stacking technique \citep[e.g.][]{carilli08, ivison10} or dedicated samples of specific objects \citep[e.g. sub-mm galaxies (SMGs);][]{kovacs06, hainline09, michalowski09}. Evolutionary studies, all based on samples poorly matched in terms of bolometric luminosity at low and high redshift, have provided conflicting results, concluding that the local IR-radio relation either does (e.g. \citealp{garrett02, appleton04, ibar08, garn09}; S10) or does not \citep[e.g.][]{seymour09, ivison10} hold out to high redshift.

\noindent Recently, predictions have been made for the redshift evolution of the IR/radio properties of star-forming galaxies having different luminosities and geometries \citep[e.g. compact starbursts and normal star-forming disks;][]{murphy09b, lackithompson09}. The current generation of IR and radio observatories can directly detect the brightest of these systems over a significant fraction of Hubble time, provided that sufficiently large cosmological volumes are sampled. Here we make use of the VLA and {\it Spitzer} coverage of the 2 deg$^2$ COSMOS field to construct (cf. \S\ref{sect:data}) a volume-limited sample of ULIRGs at $z<$\,2 that allows a direct comparison of observations and theory. Our findings are presented in \S\ref{sect:results} and discussed in \S\,\ref{sect:conclusions}.

\noindent We adopt the WMAP-5 cosmology \citep[$\Omega_m$\,=\,0.258, $\Omega_{\Lambda}$\,+\,$\Omega_m$\,=\,1 and $H_0$\,=\,71.9 km\,s$^{-1}$\,Mpc$^{-1}$;][]{dunkley09}.

\section{Data and Sample Selection}
\label{sect:data}

\subsection{IR and Radio Measurements}
\label{sect:IRradiodat}

The 1.4\,GHz map \citep{vlacos2} of the 2\,deg$^2$ COSMOS field reaches an average sensitivity of $\sim$0.017\,mJy/beam (FWHM\,=\,2.5$''$). Here we use the VLA-COSMOS `Joint' Catalog (Schinnerer et al. 2010, subm.) containing $\sim$2900 sources detected with $S/N\geq$ 5. {\it Spitzer}/MIPS imaging by the S-COSMOS project \citep{sanders07} achieves a resolution of 5.8$''$ (18.6$''$) and a 1\,$\sigma$ point source detection limit of $\sim$0.018 (1.7)\,mJy at 24 (70)\,$\mu$m (for details see \citealp{lefloch09} and \citealp{frayer09}). The depth of the 24\,$\mu$m observations exceeds that at 70\,$\mu$m and 1.4\,GHz by roughly a factor of seven in terms of equivalent IR luminosity (cf. Fig. 1 in S10). At an equal detection significance level (3\,$\sigma$), the 24\,$\mu$m catalog consequently is roughly 20-fold larger than the 70\,$\mu$m source list ($\sim$50,000 vs. 2,700). {\it Spitzer} detections were matched to the VLA-COSMOS sources using a search radius of $\nicefrac{\rm FWHM}{3}$ for either IR filter. Ambiguous radio-IR associations were removed from the sample. Since the 1.4\,GHz catalog is restricted to sources with $S/N$\,$\geq$\,5, we re-analyzed the radio map at the position of unmatched IR sources and added all resulting detections with $S/N$\,$>$\,3 ($\sim$2100 objects) to the sample. For more information on the band-merging and the flux distributions of the (matched and unmatched) IR and radio sources we refer to the detailed description in S10.\\

\noindent We use the joint flux information at 24 and 70\,$\mu$m to determine -- given the known redshift (cf. \S\,\ref{sect:znclass}) -- the best-fitting synthetic IR spectral energy distribution (SED) and thence the IR luminosity, $L_{\rm TIR}$\,$\equiv$\,$L(8-1000\,\mu{\rm m})$. As described in S10 \citep[\S\,4; see also][for additional details on the SED fitting]{murphy09a}, templates according to \citet{charyelbaz01} are used for galaxies directly detected in both MIPS filters. For sources only detected at 24\,$\mu$m we also fit the {\it Spitzer} photometry (including the 70\,$\mu$m upper flux limit) with \citet{dalehelou02} template SEDs  and define the best estimate of the IR luminosity as the average $L_{\rm TIR}$ from the two separate fits. Inferring $L_{\rm TIR}$ from only two bands at wavelengths shorter than the peak of the IR emission is expected to lead to uncertainties of a factor two to five \citep{murphy09a, kartaltepe10}. Although our estimates of $L_{\rm TIR}$ are thus less precise than those presented by, e.g., \citet{ivison10} for a similar study, there is no consensus in the literature that they are systematically biased to low or high fluxes (cf. discussion in S10, \S\,6.5).

\noindent The IR/radio properties of our sources are characterized by the logarithmic TIR/radio flux ratio $q_{\rm TIR}$ \citep{helou85}:
\begin{equation}
q_{\rm TIR} = {\rm log}\left(\frac{L_{\rm TIR}}{3.75\times10^{12}\,{\rm W}}\right) - {\rm log}\left(\frac{L_{\rm 1.4\,GHz}}{\rm W\,Hz^{-1}}\right)~.
\label{eq:qTIR}
\end{equation}
The rest-frame 1.4\,GHz luminosity $L_{\rm 1.4\,GHz}$ is
\begin{equation}
L_{\rm 1.4\,GHz} {\rm [W\,Hz^{-1}]} = \frac{4\pi D_L(z)^2}{(1+z)^{1-\alpha}}\,S_{\nu}({\rm 1.4\,GHz})~,
\label{eq:L20}
\end{equation}
where $S_{\nu}({\rm 1.4\,GHz})$ is the observed integrated radio flux density of the source and $D_L(z)$ the luminosity distance. The $K$-correction $(1+z)^{-(1-\alpha)}$ depends on the spectral index of the synchrotron emission, which is set to $\alpha=$ 0.8 \citep{condon92}. Given the mean spectral slopes typically measured for faint extragalactic radio sources \citep[0.4\,$\lesssim$\,$\alpha$\,$\lesssim$\,0.9; e.g.][]{ibar09} our values of $L_{\rm 1.4\,GHz}$ should be accurate to within 40 (70)\% at $z$\,$\sim$\,1 (2). The main contribution to uncertainties on $q_{\rm TIR}$ thus stems from errors on $L_{\rm TIR}$.

\subsection{Distances and Source Classification}
\label{sect:znclass}

Optical data and photometric redshifts\footnote{~The photo-$z$ dispersion $\sigma(\nicefrac{\Delta z}{(1+z)})$ is 0.007, 0.013 and 0.051 for sources at $z\,<$\,1.25 with $i_{\rm AB}$\,$<$\,22.5, $i_{\rm AB}$\,$\in$\,[22.5,\,24] and $i_{\rm AB}$\,$>$\,24, respectively. At higher redshifts the accuracy of the photometric redshifts decreases by a factor of $\sim$3.} are taken from the catalog of \cite{ilbert09}. The wavelength range spanned by these observations (30 broad, medium and narrow band filters) extends from 1550\,\AA\ to 8\,$\mu$m. \cite{capak07, capak08} provide a complete description of these observations. Spectroscopic redshifts from the zCOSMOS survey \citep{lilly09} or Magellan/IMACS and Keck/Deimos follow-up observations (\citealp[e.g.][]{trump09}; Kartaltepe et al., in prep.) are available for $\sim$25\% of our sources, most of which lie at $z$\,$\lesssim$\,1. (See values of f$_{\rm spectro}(z)$, the spectroscopically observed sample fraction, in Fig. 2.) The quality of distance measurements is assessed using spectroscopic confidence flags or the width of the photo-$z$ probability distribution in the case of photometric redshifts (see S10, Appendix). Sources that do not fulfill the reliability requirements are excluded from the subsequent analysis. The optical photometry and spectroscopy was matched to the radio and IR catalog entries using a search radius of 0.6$''$ and 1$''$ (reflecting the larger uncertainty on the centroids of IR sources), respectively. As done for the band-merging of the {\it Spitzer} and VLA data, ambiguous optical-IR/radio associations are discarded.

\noindent In Fig. \ref{fig:malmquist} we plot the IR luminosities of our sources as a function of their redshift. Based on the range of accessible IR luminosities at each redshift we define two populations for later investigation: ({\it i}\,) ULIRGs with $L_{\rm TIR}$\,$\geq$\,10$^{12}$\,$L_{\sun}$, and ({\it ii}\,) all objects populating the bright end of the TIR luminosity function (LF) derived by \citet{magnelli09}. The bright end is defined as $L_{\rm TIR}$\,$\geq$\,$L^{\rm (knee)}_{\rm TIR}(z)$, where $L^{\rm (knee)}_{\rm TIR}(z)$\,$\propto$\,(1+$z$)$^{3.6\pm0.4}$ represents the break in a double power law parameterization\footnote{~Consistent with the measurement of \citet{magnelli09}, \cite{lefloch05} and \cite{caputi07} report an evolution of (1+$z$)$^{3.2^{+0.7}_{-0.2}}$ and (1+$z$)$^{3.5\pm0.4}$, respectively, based on a {\it double exponential} fit to the TIR LF at $z$\,$\lesssim$\,1.} for the TIR LFs. Both selection approaches lead to a volume-limited sample of either ULIRGs or `IR-bright' galaxies spanning the range $z$\,$\lesssim$\,2.

\noindent We divide our sample into star forming (SF) galaxies and active galactic nuclei (AGN) using a modification of the rest-frame optical color-based method developed by \citet{smolcic08}. Optical--to--near-IR SED fits with the package ZEBRA \citep{feldmann06} provide rest-frame ($u-K$)-colors that are translated into the probability of `SF-hood', Pr\,(SF), for each object in our sample (cf. \S\,3 in S10). SF systems are all galaxies with Pr\,(SF)\,$>$\,0.5 (or ($u-K$)$_{\rm AB}$\,$<$\,2.36). Galaxies with redder colors are regarded as AGN hosts.

\section{Results}
\label{sect:results}

\begin{figure}
\centering
\includegraphics[scale=0.65, angle=-90]{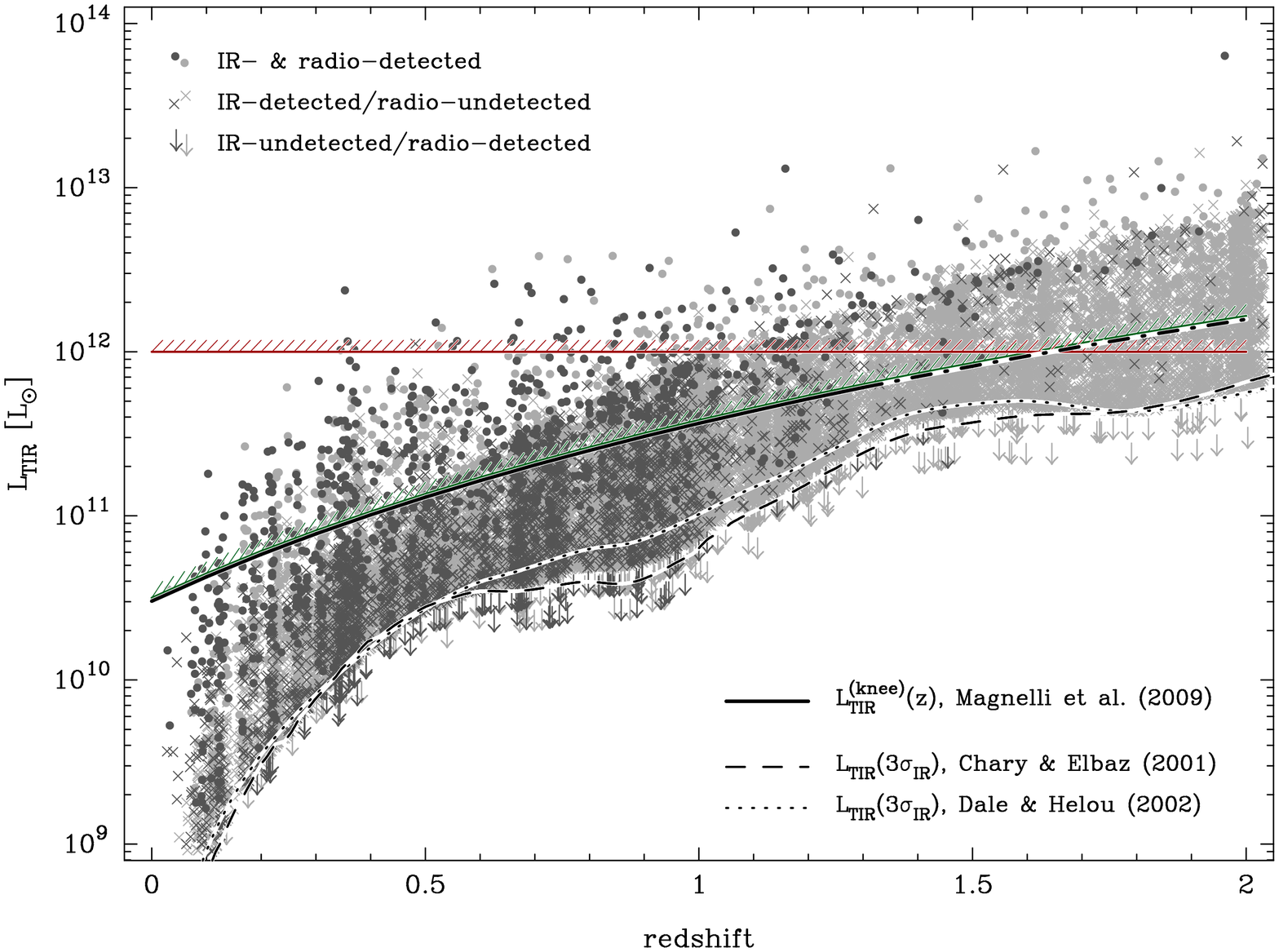}
\caption{Infrared luminosities $L_{\rm TIR}$ of our 24\,$\mu$m- and/or 1.4\,GHz-detected sources (see legend in the upper left corner) as a function of redshift. The dotted/dashed line tracing the lower edge of the measurements mark the minimal detectable $L_{\rm TIR}$ as predicted (based on 24\,$\mu$m sensitivity) by template SEDs. Dark (light) grey symbols denote sources with spectroscopic (photometric) redshifts. The solid black line shows the evolution of the characteristic luminosity $L^{\rm (knee)}_{\rm TIR}$ of the TIR LF at $z$\,$<$\,1.3 \citep{magnelli09}. Its extrapolation to $z$\,$\sim$\,2 is schematically indicated by the dash-dotted segment. Red and green lines delimit the luminosity range of the two samples defined in \S\,\ref{sect:znclass}. \label{fig:malmquist}}
\end{figure}

To constrain the evolution of average IR/radio ratios we compute the median, $\langle q_{\rm TIR}\rangle$, in different redshift slices. The selection threshold for ULIRGs and the IR-bright population lies well above the faintest accessible IR luminosities in the redshift range 0\,$<$\,$z$\,$<$\,2 (c.f. Fig. \ref{fig:malmquist}). Our samples are a mixture of ({\it i}\,) sources directly detected at IR {\it and} radio wavelengths, plus ({\it ii}\,) 24\,$\mu$m-detected sources with only a 3\,$\sigma$ upper radio flux limit from the 1.4\,GHz rms image. The corresponding IR/radio ratios are either well-defined (within experimental uncertainties) or lower limits and, when combined, form a `censored' sample that is best analyzed with the tools of survival analysis. In the present case of one-sided censoring, the cumulative distribution function (CDF) of measurements of $q_{\rm TIR}$ can be derived with the \citet{kaplanmeier58} product limit estimator. As it is normalized (i.e. runs from zero to unity), the median corresponds to that $q_{\rm TIR}$ for which the CDF is equal to 0.5.

\noindent We derive $\langle q_{\rm TIR}\rangle$ for our sample of 1,692 SF ULIRGs and for all 3,132 COSMOS ULIRGs (SF and AGN). Fig. \ref{fig:distribs_a} shows the accordingly normalized CDFs. At $q_{\rm TIR}\gg\langle q_{\rm TIR}\rangle$, the CDFs approach a non-zero value reflecting the number of lower limits on $q_{\rm TIR}$ that may exceed the largest uncensored measurement in that bin. Due to the comparatively shallow VLA observations, IR sources without a directly detected 1.4\,GHz counterpart become more frequent as redshift increases. Therefore, the width of the redshift bins chosen for the construction of the CDFs compromises between a split of the studied redshift range $z$\,$<$\,2 into regular intervals and the aim to sample the distribution function down to the median. In Fig. \ref{fig:distribs_b} we show the CDFs for IR-bright COSMOS sources (3,004/5,657 in the SF/total sample, respectively) that satisfy $L_{\rm TIR}$\,$\geq$\,$L^{\rm (knee)}_{\rm TIR}(z)$. The larger number of faint(er) luminosity sources allowed us to divide the range $z$\,$\lesssim$\,1.6 into thinner slices than was done for ULIRGs. Moreover, there was a sufficiently large number of objects of this class even at low redshift (0\,$<$\,$z$\,$<$\,0.2), whereas the closest ULIRG in our sample lies at $z$\,$\sim$\,0.4 (cf. Fig. \ref{fig:malmquist}).

\noindent The values of $\langle q_{\rm TIR}\rangle$ reported in Table \ref{tab:survmeds} increase at $z$\,$\gg$\,1. In S10 (Fig. 17 and Table 6) we found that -- although average IR/radio ratios display little evolution with redshift -- the dispersion $\sigma_{q_{\rm TIR}}$ in the COSMOS data increases fairly abruptly at $z$\,$\sim$\,1.4. This is primarily due to increased uncertainties in $q_{\rm TIR}$, but might also hide a small intrinsic increase in the dispersion. The scatter $\sigma_{q_{\rm TIR}}$ directly influences the shift \citep[e.g.][]{kellermann64}
\begin{equation}
\Delta q_{\rm bias} = {\rm ln}(10)\,(\beta-1)\,\sigma_{q_{\rm TIR}}^2
\label{eq:qbias}
\end{equation}
between the average IR/radio ratio of flux-limited samples selected at IR and radio wavelengths ($\beta$ is the power law index of the differential source counts $\nicefrac{dN}{dS} \propto S^{-\beta}$). S10 showed that eq. (\ref{eq:qbias}) predicts the actual offsets present in the COSMOS data remarkably well. Because of the higher sensitivity of the 24\,$\mu$m observations, the present sample is effectively IR-selected. Eq. (\ref{eq:qbias}) allows us to compensate for the relative offset between medians at high and low redshift that arises {\it artificially} due to the increased scatter in our data at $z$\,$\gtrsim$\,1.4. In doing so, we ({\it i}\,) use that $\overline{\sigma}_{q_{\rm TIR}}(z\lesssim1.4) \approx 0.35$ and $\overline{\sigma}_{q_{\rm TIR}}(1.4\lesssim z < 2) \approx 0.75$ (cf.  S10, Table 6), and we ({\it ii}\,) assume that the observed flux densities of galaxies at $z$\,$>$\,1.4 primarily lie in a range of sub-Euclidean source counts where $\beta$\,$\approx$\,1.5 \citep[e.g.][]{chary04, papovich04}. The resulting correction (to be subtracted from the medians at $z$\,$>$\,1.4) is
\begin{eqnarray}
{\rm D}q_{\rm corr.} &=& \Delta q_{\rm bias}(1.4\lesssim z < 2) - \Delta q_{\rm bias}(z<1.4) \\
 &=& {\rm ln}(10)\left[(1.5-1)\times0.75^2 - (2.5-1)\times0.35^2\right] \nonumber \\
 &\simeq& 0.22~, \nonumber
\label{eq:qbiasfix}
\end{eqnarray}
with an associated uncertainty (owing to the errors on $\overline{\sigma}_{q_{\rm TIR}}$ and $\beta$) of approximately 0.13. Our step function-like correction neglects that sources at a given redshift may be drawn from a flux range with continuously varying $\beta$. This is the simplest possible form that allows us to correctly compensate for an apparent, spurious offset in $\langle q_{\rm TIR}\rangle$ between the limits of our investigated redshift range.

\noindent In Fig. \ref{fig:evosummary} we plot $\langle q_{\rm TIR}\rangle$ vs. redshift and relate these medians with the best-fitting evolutionary trend of the form $\nicefrac{\langle q_{\rm TIR}(z)\rangle}{\langle q_{\rm TIR}(z=0)\rangle} \propto (1+z)^\gamma$ for each of our (sub-)samples. To constrain the fit at low redshift we add a low-$z$ data point (both for the ULIRG and the IR-bright sample) based on the complete IRAS-selected sample of \citet{yun01}. The values of $q_{\rm FIR}$ given by \citet{yun01} were converted to $q_{\rm TIR}$ by boosting their IR flux by a factor of two\footnote{~Due to the well-constrained mean IR/radio ratios in the samples of \citet{yun01} ($\langle q_{\rm FIR}\rangle$\,=\,2.34$\pm$0.01) and \citet{bell03} ($\langle q_{\rm TIR}\rangle$\,=\,2.64$\pm$0.02) this {\it average} correction factor is accurate to within a few percent.}, the average difference between the mean $q_{\rm TIR}$ and $q_{\rm FIR}$ found by \citet{bell03}. Since SF systems in the sample of \citet{yun01} could not be identified following the procedure employed for the COSMOS galaxies (see \S\,\ref{sect:znclass}), our reference sample for SF galaxies simply consists of all local sources with $L_{\rm 1.4\,GHz}$\,$<$\,$10^{24}$\,W/Hz, in keeping with \citet{condon89}.

\noindent The two panels of Fig. \ref{fig:evosummary} show the evolutionary trends for SF systems ({\it top}) and all galaxies (SF and AGN; {\it bottom}). To account for the asymmetric error bars, we have drawn 1001 values from within the 95\% confidence region of each median\footnote{~When upper confidence limits are $\infty$, we set the upper error bar to twice the lower one. This eases the calculation of $\gamma$, while reflecting that upper and lower confidence interval are generally similar down to the 60$^{\rm th}$ percentile of the CDF.} and then fit the evolutionary trend 1001 times based on random combinations (without replacement) of the resampled medians.  The final best-fit values given below are the medians of the parameter distributions thus obtained.\\
If the uncorrected high-$z$ medians are included in the evolutionary fit, we find an exponent $\gamma$\,=\,$0.09_{-0.07}^{+0.08}$/$0.11_{-0.05}^{+0.06}$ for ULIRGs (SF and total sample, respectively) and $\gamma$\,=\,$0.13\pm0.06$/$0.12\pm0.04$ for the IR-bright galaxies. (Errors delimit the 95\% confidence interval.) This would imply a doubling of the average TIR/radio flux ratio from redshift 0 to 2, with most of the evolution happening at $z$\,$\gtrsim$\,1.4. By taking the corrected medians at $z$\,$\gtrsim$\,1.4 into consideration, we find an evolution of the average IR/radio ratios of ULIRGs according to $(1+z)^{-0.01\pm0.06}$ (identical for SF and all ULIRGs). Similarly, the trend for the IR-bright population is consistent with being zero, both for the total sample ($\gamma$\,=\,$0.03_{-0.04}^{+0.05}$) and the SF subset where $\gamma$\,=\,$0.04_{-0.05}^{+0.06}$.

\section{Conclusions}
\label{sect:conclusions}

\noindent We have presented the first investigation of the evolution of the IR-radio relation out to $z$\,$\sim$\,2 for a statistically significant, volume-limited sample of IR-luminous galaxies. This advance became possible thanks to two factors: ({\it i}\,) the large area and deep mid-IR coverage of the COSMOS field, and ({\it ii}\,) the inclusion of flux limits in the analysis with appropriate statistical tools.\\
At redshifts $z$\,$<$\,2 the median TIR/radio ratio of ULIRGs remains unchanged if we compensate for biases. On the most basic level this implies that their magnetic fields, $B$, are sufficiently strong to ensure that cosmic ray electrons predominantly lose their energy through synchrotron radiation (rather than inverse Compton scattering off the CMB). Regarding the 0.3\,dex increase of the uncorrected evolutionary signal as an upper limit implies that $B$\,$\gtrsim$\,30\,$\mu$G \citep[e.g.][Fig. 5]{murphy09b}, as expected for compact and strong starbursts. This conclusion applies to both SF systems and optically selected AGN hosts, consistent with the similar mean IR/radio ratios reported for these two classes of objects in S10. Our finding agrees with theoretical and numerical expectations that ULIRGs should follow the local IR-radio relation until at least $z\sim$\,2 \citep{lackithompson09, murphy09b}. Moreover, it suggests that the lower IR/radio ratios frequently reported for high-$z$ SMGs (e.g. \citealt{kovacs06, valiante07, murphy09a, michalowski09}, but see also \citealt{hainline09}) are not typical of distant ULIRGs in general.

\noindent Our complete sample of `IR-bright' galaxies -- the population that resides on the evolving bright end of the TIR LF -- links high-$z$ ULIRGs to normal IR-galaxies (log($\nicefrac{L_{\rm TIR}}{L_{\odot}})\gtrsim$\,10.5) in the local universe. The fact that the average IR/radio ratio of the latter is very similar to that of ULIRGs demonstrates that the similar IR/radio properties of existing SF samples at low and high redshift are not the fortuitous consequence of comparing objects in different luminosity ranges. While distant starbursts follow the same IR-radio relation as local sources, this has not yet been ascertained for galaxies with moderate SF rates ($\lesssim$10\,$M_{\odot}$/yr) that cannot be directly detected with current radio and far-IR facilities. The recent stacking analyses of \citet{seymour09} and \citet{ivison10} measured a steady decline of average IR/radio ratios that begins at $z$\,$<$\,1 and continues out to $z$\,$>$\,2. Given that the average IR luminosities of their image stacks are comparable to those of our `IR-bright' sample these findings are at odds with our measurements and, as shown here, cannot be ascribed to a luminosity offset between low and high redshift sources. It is to be expected that the origin of the discrepancy -- possibly the different methodology, sample selection or SED evolution \citep[e.g.][]{symeonidis09, seymour10} -- will soon be identified in upcoming EVLA and Herschel surveys by virtue of the increased sensitivity and/or wavelength coverage these observatories offer. The latter capability in particular will ensure more accurate measurements of radio spectral indices and a better sampling of dust emission well into the far-IR which is crucial to the determination of the dust temperatures in distant starbursts. These improvements should also lead to a decrease in the scatter of the IR-radio relation at high redshift, thereby reducing both the need for and the impact of bias corrections of the kind that were applied in this work.

\acknowledgments MTS acknowledges DFG-funding (grant SCHI 536/3-2) and thanks Gianni Zamorani and Rob Ivison for reading and commenting on the manuscript.\\
This work is partly based on observations made with the {\it Spitzer Space Telescope}, which is operated by NASA/JPL/Caltech. The National Radio Astronomy Observatory (NRAO) is operated by Associated Universities, Inc., under cooperative agreement with the National Science Foundation.
 
\noindent {\it Facilities:} NRAO (VLA), {\it Spitzer} (IRAC, MIPS), ESO (VLT), Subaru (SuprimeCam)

\clearpage

\begin{deluxetable}{lllc@{\hspace{5truemm}}|@{\hspace{5truemm}}llc}
\tabletypesize{\scriptsize}
\tablecaption{Medians of the CDFs shown in Fig. 2. Bracketed values of $\langle q_{\rm TIR}\rangle$ are the medians in high redshift bins before correction according to eq. (\ref{eq:qbiasfix}). \label{tab:survmeds}}
\tablewidth{0pt}
\setlength{\tabcolsep}{0.06in}
\tablehead{
\colhead{sample/galaxy type} &
\colhead{$\langle z\rangle$} &
\colhead{$\langle q_{\rm TIR}\rangle$} & 
\colhead{$N$} &
\colhead{$\langle z\rangle$} &
\colhead{$\langle q_{\rm TIR}\rangle$} &
\colhead{$N$} }
\startdata
 & \multicolumn{3}{l}{\bf ULIRG sample} & \multicolumn{3}{l}{\bf `IR-bright' sample}\\[1ex]
\hline
\hline\\[-2ex]
local (all sources) & 0.099$_{-0.004}^{+0.020}$ & $2.674_{-0.079}^{+0.020}$ & 51 & 0.022$\pm$0.001 & $2.621_{-0.013}^{+0.016}$ & 1,109\\[1ex]
\hline\\[-2ex]
COSMOS & & & & 0.169$\pm$0.014 & $2.528_{-0.061}^{+0.054}$ & 34 \\[2ex]
\raisebox{1ex}[1ex]{(all sources)} & & & & 0.346$\pm$0.008 & $2.703_{-0.040}^{+0.041}$ & 231 \\[1ex]
 & & & & 0.518$\pm$0.008 & $2.629_{-0.079}^{+0.067}$ & 272 \\[1ex]
 & 0.659$\pm$0.053 & $2.622_{-0.190}^{+0.179}$ & 26 & 0.705$\pm$0.004 & $2.515_{-0.047}^{+0.017}$ & 486 \\[1ex]
 & & & & 0.906$\pm$0.005 & $2.552_{-0.034}^{+0.053}$ & 754 \\[1ex]
 & \raisebox{1.8ex}[1.8ex]{0.943$\pm$0.022} & \raisebox{1.8ex}[1.8ex]{$2.439_{-0.154}^{+0.124}$} & 52 & 1.108$\pm$0.006 & $2.612_{-0.086}^{+0.034}$ & 637 \\[1ex]
 &  \raisebox{1.8ex}[1.8ex]{1.200$\pm$0.009} &  \raisebox{1.8ex}[1.8ex]{$2.548_{-0.070}^{+0.083}$} & 213 & 1.280$\pm$0.007 & $2.619_{-0.056}^{+0.170}$ & 487 \\[1ex]
 & 1.466$\pm$0.004 & $2.520_{-0.042}^{+\infty}$ (2.740) & 715 & 1.498$\pm$0.005 & $2.576_{-0.056}^{+0.239}$ (2.796) & 1506 \\[1ex]
 & 1.628$\pm$0.003 & $2.585_{-0.057}^{+0.245}$ (2.805) & 553 & & & \\[1ex]
 & 1.785$\pm$0.004 & $2.715_{-0.144}^{+\infty}$ (2.935) & 627 & 1.784$\pm$0.004 & $2.715_{-0.146}^{+\infty}$ (2.935) & 546 \\[1ex]
 & 1.981$\pm$0.003 & $2.651_{-0.104}^{+0.105}$ (2.871) & 1,185 & 1.980$\pm$0.003 & $2.651_{-0.102}^{+0.140}$ (2.871) & 875 \\[1ex]
\hline
\hline\\[-2ex]
local (SF sources) & 0.095$_{-0.006}^{+0.015}$ & $2.703_{-0.050}^{+0.055}$ & 47 & 0.022$\pm$0.001 & $2.622_{-0.012}^{+0.017}$ & 1,101\\[1ex]
\hline\\[-2ex]
COSMOS & & & & 0.169$\pm$0.015 & $2.538_{-0.115}^{+0.141}$ & 19 \\[1ex]
\raisebox{1ex}[1ex]{(SF sources)} & & & & 0.340$\pm$0.014 & $2.716_{-0.064}^{+0.075}$ & 101 \\[1ex]
 & & & & 0.528$\pm$0.010 & $2.649_{-0.117}^{+0.086}$ & 142 \\[1ex]
 & \raisebox{1.8ex}[1.8ex]{0.624$\pm$0.062} & \raisebox{1.8ex}[1.8ex]{$2.722_{-0.279}^{+0.349}$} & 10 & 0.705$\pm$0.006 & $2.540_{-0.035}^{+0.092}$ & 242\\[1ex]
 & 0.940$\pm$0.030 & $2.439_{-0.212}^{+0.156}$ & 27 & 0.915$\pm$0.007 & $2.560_{-0.037}^{+0.053}$ & 394 \\[1ex]
 & & & & 1.107$\pm$0.008 & $2.672_{-0.061}^{+0.069}$ & 327 \\[1ex]
 & \raisebox{1.8ex}[1.8ex]{1.210$\pm$0.012} & \raisebox{1.8ex}[1.8ex]{$2.617_{-0.082}^{+0.132}$} & 119 & 1.273$\pm$0.009 & $2.709_{-0.147}^{+\infty}$ & 263 \\[1ex]
 & 1.467$\pm$0.004 & $2.534_{-0.038}^{+\infty}$ (2.754) & 412 & 1.501$\pm$0.006 & $2.584_{-0.066}^{+0.258}$ (2.576) & 932 \\[1ex]
 & 1.628$\pm$0.004 & $2.576_{-0.080}^{+0.254}$ (2.796) & 348 & & & \\[1ex]
 & 1.786$\pm$0.006 & $2.715_{-0.131}^{+\infty}$ (2.935) & 321 & 1.786$\pm$0.006 & $2.715_{-0.132}^{+\infty}$ (2.935) & 268 \\[1ex]
 & 1.984$\pm$0.004 & $2.672_{-0.121}^{+\infty}$ (2.892) & 591 & 1.982$\pm$0.005 & $2.712_{-0.160}^{+\infty}$ (2.932) & 403 
\enddata
\tablecomments{Columns are: sample description (1), median redshift (with 1\,$\sigma$ error) of redshift slice in which CDF is constructed (2; 5), median TIR/radio ratio with upper and lower 95\% confidence intervals (3; 6), number of objects used (4; 7).}
\end{deluxetable}

\clearpage

\clearpage

\newcounter{subfigure}
\renewcommand{\thefigure}{\arabic{figure}\alph{subfigure}}

\setcounter{subfigure}{1}
\begin{figure}
\centering
\includegraphics[scale=0.8]{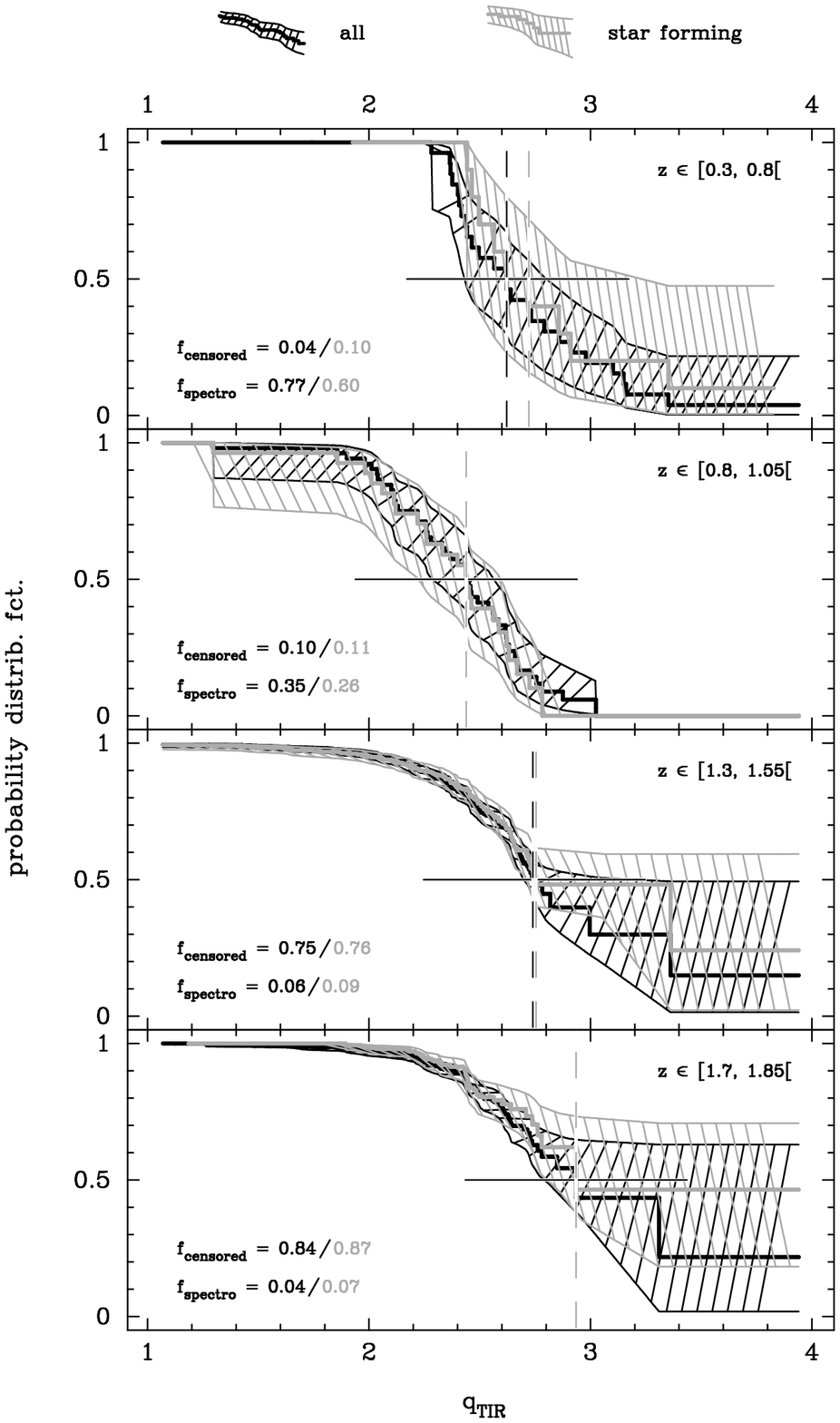}
\caption{Cumulative distribution functions (CDFs) of ULIRG TIR/radio ratios, specifying the fraction of the population with a $q_{\rm TIR}$ larger than a given value on the horizontal axis. The panels show every second redshift bin from Fig. \ref{fig:evosummary}. \newline
Hatched areas: 95\% confidence interval. Grey: CDFs of SF objects. Black: all active galaxies. The intersection of the curves with the solid horizontal line defines the median $\langle q_{\rm TIR}\rangle$ (vertical dashed lines). f$_{\rm spectro}$ specifies the sample fraction with spectroscopically measured redshifts; f$_{\rm censored}$ is the fraction for which $q_{\rm TIR}$ is constrained by a lower limit. \label{fig:distribs_a}}
\end{figure}

\clearpage

\addtocounter{figure}{-1}
\addtocounter{subfigure}{1}
\begin{figure}
\centering
\includegraphics[scale=0.8]{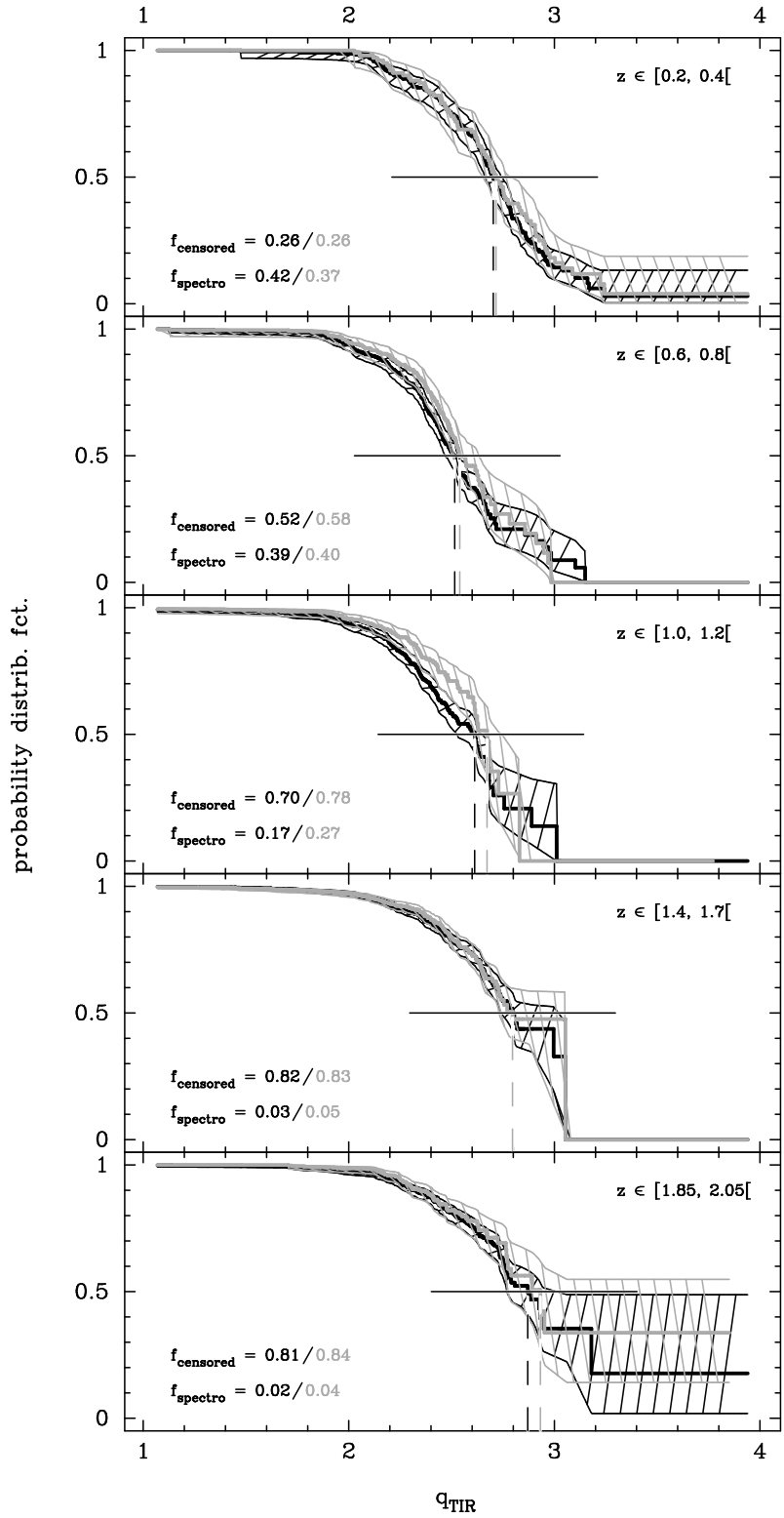}
\caption{As Fig. \ref{fig:distribs_a} but for the IR-bright population.\label{fig:distribs_b}}
\end{figure}

\clearpage

\renewcommand{\thefigure}{\arabic{figure}}

\begin{figure}
\centering
\includegraphics[scale=0.7, angle=-90]{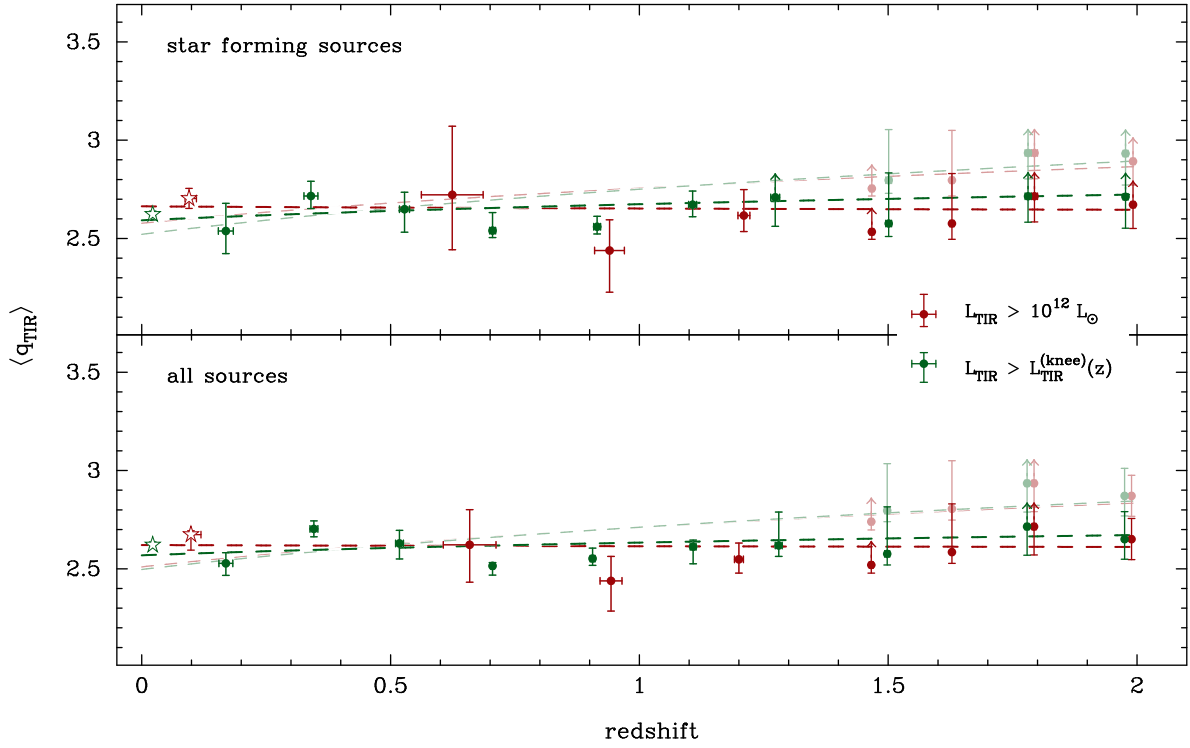}
\caption{Redshift evolution of the median logarithmic TIR/radio ratio $\langle q_{\rm TIR}\rangle$ for IR-bright galaxies ($L_{\rm TIR}$\,$>$\,$L^{\rm (knee)}_{\rm TIR}$; green symbols) and ULIRGs (red). In the upper panel we consider the subset of SF sources, extracted from the entire sample of active galaxies (bottom). Transparent symbols: estimates of $\langle q_{\rm TIR}\rangle$ prior to correction for selection biases (see \S\,\ref{sect:results}). The best-fitting evolutionary trends to the corrected (uncorrected) measurements of $\langle q_{\rm TIR}\rangle$ are reported using strong (transparent) dashed lines. They have been additionally constrained (open stars) at low redshift by the sample of \citet{yun01}.\newline
Both ULIRGs and IR-bright galaxies have constant average IR/radio properties out to $z$\,$\sim$\,2 when correcting for bias, otherwise $\sim$0.3 dex of positive evolution is found. \label{fig:evosummary}}
\end{figure}

\end{document}